\def\ln{\ell{n}}
\documentstyle[a4,12pt]{article}
\begin{document}
\begin{titlepage} \vspace{0.2in} \begin{flushright}
MITH-97/1 \\ \end{flushright} \vspace*{1.5cm}
\begin{center} {\LARGE \bf  On the Ground State of Quantum Gravity
\\} \vspace*{0.8cm}
{\bf S.~Cacciatori, G.~Preparata, S.~Rovelli, I.~Spagnolatti,
S.-S.~Xue$^{(a)}$}\\ \vspace*{1cm}
Dipartimento di Fisica dell'Universit\`a and INFN - sezione di Milano, 
Via Celoria 16, Milan, Italy\\
(a)I.C.R.A.-International Center for Relativistic Astrophysics, La
Saoienza 00185 Rome, Italy \\ \vspace*{1.8cm}

{\bf   Abstract  \\ } \end{center} \indent
 
In order to gain insight into the possible Ground State of Quantized Einstein's
Gravity, we have devised a variational calculation of the energy of the quantum
gravitational field in an open space, as measured by an asymptotic observer
living in an asymptotically flat space-time. We find that for Quantum Gravity
(QG) it is energetically favourable to perform its quantum fluctuations not
upon flat space-time but around a ``gas'' of wormholes, whose size is the Planck
length $a_p$ ($a_p\simeq 10^{-33}$cm). As a result, assuming such configuration
to be a good approximation to the true Ground State of Quantum Gravity,
space-time, the arena of physical reality, turns out to be well described by
Wheeler's Quantum Foam and adequately modeled by a space-time lattice with
lattice constant $a_p$, the Planck lattice. 

\vfill \begin{flushleft}  January, 1997   \vspace*{3cm} \\
\noindent{\rule[-.3cm]{5cm}{.02cm}} \\
\vspace*{0.2cm} \end{flushleft} \end{titlepage}
 
\section{\bf Introduction}
 
Among the fundamental interactions of Nature, since the monumental contribution
of Albert Einstein, Gravity plays the central role of determining the structure
of space-time, the arena of physical reality. As well known, in classical
physics a world without matter, the Vacuum, has the simplest of all structures,
it is flat (pseudoeuclidean); but in quantum physics? This is the central
question that has occupied the best theoretical minds since it became
apparent, at the beginning of the 30's, that Quantum Field Theory (QFT) is
the indispensable intellectual tool for discovering the extremely subtle ways
in which the quantum world actually works. Thus the problem to solve was to
find in some way or other the Ground State (GS) of Quantum Gravity (QG), which
determines the dynamical behaviour of any physical system, through the
non-trivial structure that space-time acquires as a result of the quantum
fluctuations that in such state the gravitational field, like all quantum
fields, must experience. Of course this problem, at least in the
non-perturbative regime, is a formidable one, and many physicists, J.A.~Wheeler
foremost among them, could but speculate about the ways in which the expected
violent quantum fluctuations at the Planck distance $a_p$ ($a_p\simeq 
10^{-33}cm$) could change the space-time structure of the Vacuum 
from its classical, trivial (pseudoeuclidean) one. And Wheeler's 
conjecture \cite{w}, most imaginative and
intriguing, of a space-time foam vividly expresses the intuition that at the
Planck distance the fluctuations of the true QG ground state would end up in
submitting the classical continuum of events to a metamorphosis into an
essentially discontinuous, discrete structure.\footnote{We should like to recall
here that, based on Wheeler's idea, a successful research program was initiated
a few years ago to explore the consequences of the Standard Model
($SU_c(3)\otimes SU_L(2)\otimes U_Y(1)$) in a discrete space-time, conveniently
modeled by a lattice of constant $a_p$, the Planck lattice (PL). For a recent 
review one may consult ref.\cite{gpx}.} 

It is the purpose of this letter to report on the recent results of an
investigation on a possible QG ground state. The starting point of our attack
is the realization that QG can be looked at as a non-abelian gauge theory whose
gauge group is the Poincar\'e group. Following the analysis performed by one of
us (GP) \cite{gp} of another non-abelian gauge theory QCD (whose gauge group is
$SU_c(3)$), we decided to explore the possibility that the energy density (to
be appropriately defined, see below) of the quantum fluctuations of the
gravitational field around a non-trivial classical solution of Einstein's
field equations for the matterless world, could be lower than the energy of the
perturbative ground state (PGS), which comprises the zero point fluctuations of
the gravitational field's modes around flat space-time. Indeed in QCD it was
found that the unstable modes (imaginary frequencies) of the gauge fields
around the classical constant chromomagnetic field, solution of the empty space
Yang-Mills equations, in the average screen completely the classical
chromomagnetic field, allowing the interaction energy between such field and
the short wave-length fluctuations of the quantized gauge field to lower the
energy density of such configuration below the PGS energy density. Thus we
decided to try for QG the strategy that was successful in QCD, i.e. 

\begin{enumerate} 

\item select a class of empty space classical solutions
of Einstein's equations that is simple and manageable; 

\item evaluate the spectrum of the small amplitude fluctuations of the 
gravitational field around such solutions; 

\item set up a variational calculation of the appropriately defined
energy density in the selected background fields; 

\item study the possible screening of the unstable modes (if any) of the
classical background fields. 

\end{enumerate} 

As for point (1) we have chosen Schwarzschild's
wormhole-solutions \cite{s}, the simplest class of solutions of Einstein's
equations after flat space-time. In order to achieve (2) the
Regge-Wheeler \cite{rw} expansion has been systematically employed, yielding two
well defined sets of {\it unstable modes} (for S- and P-wave). This important
result, already indicated in previous independent work \cite{d}, renders
the development of the points (3) and (4) both relevant and meaningful, the
former point yielding a lowering of the energy density due to the interaction 
of the short-wave length
modes with the background gravitational field, the latter exhibiting the
(approximate) cancellation of the independent components of the Riemann 
tensor of 
Schwarzschild's wormholes by the S-wave unstable mode. As a result flat
space-time, like the QCD perturbative ground state, becomes ``essentially
unstable'', in the sense that upon it no {\it stable} quantum dynamics can be
realized. On the other hand a well defined ``gas of wormholes'' appears as a
very good candidate for the semi-classical configuration 
around which the quantized
modes of the gravitational field can {\it stably} fluctuate. But a discussion
of the physics implications of our findings must await a more detailed
description of our work, which we are now going to provide. 

\section{\bf The Schr\"odinger functional approach}

In order to develop a functional strategy aiming to determine the
Ground State of Quantum Gravity, which parallels the approach developed for
QCD \cite{gp}, we must first identify an appropriate energy functional. 
In General Relativity, this is a non-trivial problem for, as is well known, in
the canonical quantization procedure, first envisaged by Dirac \cite{dirac} and
Arnowitt,
Deser and Misner (ADM) \cite{a}, due to general covariance
the local Hamiltonian is constrained to annihilate 
the physical ground state, a fact
that in the Schr\"odinger functional approach is expressed by the
celebrated Wheeler-DeWitt equation \cite{dewitt}. However we note that 
the problem we wish to solve concerns 
the minimization of the total energy of an ``open space'', in
which there exists a background metric field that becomes 
``asymptotically flat'', i.e.~that for spatial infinity ($|\vec x|\rightarrow
\infty$) behaves as 
\begin{equation}
g_{ij}\rightarrow \delta_{ij}+O({1\over r}).
\label{asy}
\end{equation}
Thus, we shall have to consider the ADM-energy \cite{a}, which 
in cartesian coordinates is given by ($\partial\Sigma$ is the boundary of the 
space-region $\Sigma$, ``$,k$'' denotes partial derivative 
with respect to $x_k$)
\begin{equation}
E_{ADM}={1\over16\pi G}\int_{\partial\Sigma}dS^k \delta^{ij}
(g_{ik,j}-g_{ij,k}),
\label{adm}
\end{equation}
where
\begin{equation}
g_{ij}(x)=\eta_{ij}(x)+h_{ij}(x),
\label{h}
\end{equation}
$\eta_{ij}(x)$ being the ``asymptotically flat'' (see condition (\ref{asy}))
background field, solution of the classical vacuum Einstein's equations.
We should like to point out that $E_{ADM}$ is just the energy that an 
asymptotic observer attributes to space whose time foliation he is keeping 
anchored to his (asymptotically) flat metric.

We can now expand the total ( the sum of the Hamiltonian and $E_{ADM}$) energy
$E$ of space in powers of the quantized fluctuations $h_{ij}(x)$:
\begin{equation}
E=E^{(0)}_{ADM}+E^{(1)}_{ADM}+\sum_n\int_\Sigma d^3x (NH+N_iH^i)^{(n)},
\label{adm1}
\end{equation}
where $H$ and $H^{i}$ are the super-hamiltonian 
and super-momentum operators, as defined by ADM \cite{a}, and $N$ and $N_i$
are the ``lapse-function''
and the ``shift-vector'' of the foliation of the 4-dimensional metric 
$g_{\mu\nu}$ \cite{a} respectively.

One can easily show that, for a 
static background\footnote{A detailed account of all the calculations of
this letter will be published elsewhere \cite{will}.}
\begin{equation}
E^{(1)}_{ADM}={1\over16\pi G}\int_{\partial\Sigma}dS^k (h_{kj,j}-h_{jj,k})
=-\int_\Sigma d^3x (NH+N_iH^{i})^{(1)},
\label{adm1'}
\end{equation}
thus we can rewrite (\ref{adm}) as
\begin{equation}
E=E^{(0)}_{ADM}+\sum_{n\ge2}\int_\Sigma d^3x (NH+N_iH^{i})^{(n)},
\label{adm2}
\end{equation}
The evolution of the physical states 
$\Psi[h_{ij}]$ with respect to the fixed time of the asymptotic observer 
is governed by the ``Schr\"odinger equation"
\begin{equation}
i\hbar \partial_t\Psi[h_{ij}]=E\left[ h_{kl},-i\hbar {\delta\over
\delta h_{kl}}\right]\Psi[h_{ij}],
\label{se}
\end{equation}
and it is in this equation, whose Hamiltonian operator is given by (\ref{adm2})
and whose wave-functionals obey (order by order in $h_{ij}$) 
the superhamiltonian and supermomenta constraints \cite{dirac, a}, that 
the parallelism of our problem
with the QCD problem \cite{gp} is fully regained.
Thus in the following we shall look for the minimization of the energy
$E^{(2)}$ ($E^{(2)}$ denotes (\ref{adm2}) truncated at $n=2$):
\begin{equation}
E^{(2)}=\int[dh_{ij}]\Psi^*[h_{ij}]E^{(2)}\left[ h_{kl},-i\hbar {\delta\over
\delta h_{kl}}\right]\Psi[h_{ij}],
\label{e2}
\end{equation}
on a class of gaussian wave-functionals $\Psi[h_{ij}]$, whose arguments $h_{ij}$
fluctuate around the ``wormhole solution'' 
discovered by Schwarzschild in 1916 \cite{dirac}, 
whose line elements in polar coordinates are given
by ($2GM < r < +\infty$)
\begin{equation}
ds^2=-{r-2GM\over r} dt^2 +{r\over r-2GM}dr^2
+r^2(d\theta^2+\sin^2\theta d\phi^2)
\label{sm}
\end{equation}
and depend on the single parameter $M$, the ADM-mass, such that 
\begin{equation}
E^{(0)}_{ADM}=M.
\label{adm3}
\end{equation}

\section{\bf The unstable modes around a single 
``wormhole'': the instability of the Perturbative Ground State}

A crucial step in the solution of our problem is the diagonalization of
the operator $O^{ijkl}$, defined by $(N^{(0)}=\sqrt{1-{2MG\over r}})$
\begin{equation}
\int_x N^{(0)} V^{(2)}=
{1\over16\pi G}\int_x {1\over4N^{(0)}} h_{ij}O^{ijkl} h_{kl},
\label{n}
\end{equation}
where $(\eta = \det \eta_{ij}$, $\eta_{ij}$ being the spatial Schwarzschild
metric)
\begin{equation}
\int_x=\int d^3x\sqrt{\eta},
\label{eta}
\end{equation}
and 
\begin{equation}
V^{(2)}=-{1\over16\pi G}(R^{(2)}+{1\over2}h^k_kR^{(1)}),
\label{v2}
\end{equation}
$R^{(1)},R^{(2)}$ being the first and second order expansions in $h_{ij}$
of the scalar of curvature. Such diagonalization has to be carried out in 
the space of the wave-functions $h_{ij}(\vec{x})$, subject to the constraints 
(``$|$'' denotes the covariant derivative with respect to the background
metric)
\begin{equation}
h^k_k=0,
\label{hkk}
\end{equation}
and 
\begin{equation}
\left({h^{ij}\over N^{(0)}}\right)_{|j}=0,
\label{hkk'}
\end{equation}
as demanded by the general covariance.

Calling $\lambda(\rho)$ the eigenvalues of $O^{ijkl}$ ($\rho$ being
a complete set of labels), the variational calculation on the gaussian 
wave-functional \cite{will} shows that the total energy of the system
is given by
\begin{equation}
E=M+{\hbar \over2}\sum_\rho\sqrt{\lambda(\rho)},
\label{et}
\end{equation}
which implies that the PGS, corresponding to $M=0$, is stable only if 
all eigenvalues $\lambda(\rho)>0$. Thus if we find some field modes for 
which $\lambda(\rho)<0$, to be called ``unstable modes'', the implication 
is that
the simple minimization (\ref{et}) of the second order operator is invalid,
requiring the consideration of higher order terms \cite{gp}: at present a 
formidable undertaking. However, the discovery of negative eigenvalues is
the unmistakable sign that the PGS is unstable and that
Perturbation Theory (PT) is a totally misleading 
approach to the calculation of the
quantum corrections to the free theory (in our case flat space-time). In view of
the severe unrenormalizability of the PT of Quantum Gravity this
latter fact is certainly good news. Also, as noticed 
in Ref.\cite{gp}, the presence 
of ``unstable modes'' implies
that their contribution to the total energy $E$ is not $O(\hbar)$, 
as for the ``stable modes'', but is $O(1)$: a
classical contribution, like the ADM-mass $M$. This means that beyond the 
quadratic approximation the ``classical'' contribution from the unstable sector
of the $h_{ij}$-modes will ``screen'' the ADM-mass by a term $-\epsilon M$
($0<\epsilon <1$). Below we shall argue that the screening is actually complete
(i.e.~$\epsilon=1$) like it happens in QCD \cite{gp, gp1}.

A careful analysis \cite{will} of the eigenvalue problem
posed by the operator $O^{ijkl}$ in the space of eigenmodes obeying (\ref{hkk},
\ref{hkk'}) reveals that around a single ``wormhole'' there exist at least four
unstable modes, one in $S$-wave and three 
in P-wave with degenerate eigenvalue
\begin{equation}
\lambda=-{1\over16(GM)^2},
\label{ev1}
\end{equation}
The contribution of the stable modes ($\lambda(\rho)$ positive) to the energy 
(\ref{et}) can be easily evaluated through a WKB-analysis \cite{will}, that 
shows it to be smaller than the zero-point 
graviton contribution to the PGS energy.
This important result is due to the red-shift experienced by these modes in the 
gravitational field of the ``wormhole''. Defining,
\begin{equation}
\Delta E(M)=E(M)-E(0),
\label{de}
\end{equation}
$E(0)$ being the PGS energy in the quadratic approximation, we can finally
write ($\Lambda$ is the ultraviolet cutoff)
\begin{equation}
\Delta E(M)=M(1-\epsilon)-{64\Lambda^4\over\pi^3}R^2GM\ln\left({R\over2GM}
\right),
\label{ev1'}
\end{equation}
where $R$ ($R\gg 2GM$) is the radius of the spherical region that surrounds the
``wormhole'' where the space metric differs appreciably from the metric of the
asymptotic observer. From (\ref{de}) one can obtain an ``average'' energy
density through simple division by the volume ${4\over3}\pi R^3$ of the
spherical region. As expected the quantum corrections to the classical
``wormhole'' energy $M(1-\epsilon)$ (which comprises the screening of 
the ``unstable modes'') turns out to be negative. 

Following a general idea due to R.~Feynman \cite{gp1} we shall now argue that
$\epsilon=1$, i.e.~the ``screening'' of the unstable modes is total. Indeed,
from the explicit solutions of the ``unstable modes'' we may compute their
contribution to the Riemann tensor up to a possible multiplicative factor, $A$,
the square of their (unknown) amplitudes \cite{will}. Averaging the components
of the classical part of the full Riemann tensor over the spherical region of
radius $R$, we find an approximately vanishing value for the ``average''
classical Riemann tensor, implying that the asymptotic
observer will perceive the classical configuration of the metric field inside
the spherical region as (approximately) flat. 

In this way flat space-time is seen to be quantum mechanically unstable, the
energy density of space around a wormhole of extension $r_{WH}=2GM$
differring from it by the quantity
\begin{equation}
\Delta E(M,R)\simeq 
-{48\over\pi^4}\Lambda^4{GM\over R}\ln\left({R\over2GM}\right).
\label{de'1}
\end{equation}
 The question now is what is the ``multi-wormhole'' configuration
that minimizes the energy density of the whole space? We shall try to give a 
plausible answer to this question in the next Section.

\section{\bf The ``wormhole gas'', the possible GS of QG and the 
Planck lattice}

The analysis we have carried out so far has yielded the expression (\ref{de'1})
for the average energy density gain (with respect to flat space-time) of a
spherical region of radius $R$ around a ``wormhole'' of size $2GM$. In order to
minimize the energy of the quantum gravitational field in the whole space it
seems natural to consider a collection of such regions filling space
completely. We are thus led to look at a ``gas of wormholes'' with density
$\sim {1\over R^3}$ and size $2GM\le R$ as the (quasi)-classical field
configuration around which the gravitational quantum field fluctuates: our
model for the Ground State of Quantum Gravity. 

It is important to note that a ``gas of wormholes'' cannot form a stable 
background field for the quantized gravitational field unless the minimum 
distance between any two of them, due to 
their gravitational interaction, is bigger than
their size $2GM$. This observation clearly implies a constraint on the ADM-mass
$M$. Let us see how. Following the interesting work of Gibbons and 
Perry \cite{gpy} we know that the interaction potential 
between two wormholes of ADM-mass $M$ is just the Newtonian one
\begin{equation}
U(d)=-{GM^2\over d},
\label{nt}
\end{equation}
$d$ being the distance of their centers. Thus in classical physics there is 
nothing to prevent their collapse into a single wormhole of appropriate 
ADM-mass $M'<2M$. This situation of basic classical instability changes
when we treat the wormholes as quantum particles of mass $M$ and size $2GM$,
as appropriate to their physical nature as perceived
by the asymptotic observer. The ensuing quantum mechanical problem, akin to
the paradigmatic hydrogen atom problem, yields (for the rest of this paper
$\hbar =c=1$)
\begin{equation}
r_\circ={2\over{GM^3}}
\label{ra}
\end{equation}
for the ``Bohr-radius'' of the quantum mechanical ground state, and
\begin{equation}
E_W=-{1\over2}{GM^2\over r_\circ}=-{G^2M^5\over4}
\label{ew}
\end{equation}
for the binding energy of the two-wormholes system. Thus in order to have a
well defined and stable gas of wormholes we must have 
\begin{equation}
r_\circ={2\over GM^3}>2GM
\label{ra1}
\end{equation}
i.e.
\begin{equation}
M<G^{-{1\over2}}=m_p,
\label{mp1}
\end{equation}
stating that the mass of the wormholes of the gas must be smaller than the 
Planck-mass $m_p$. A very important finding.

We are now in a position to determine the actual size $R$ of the spherical 
domains and of
the wormholes $GM< G^{{1\over2}}$ (the Planck length, $a_p$) of the Ground state
of Quantum Gravity. Indeed, in order to achieve this we only need minimize 
the energy density ($R=2\lambda GM$)
\begin{equation}
\Delta E(M,\lambda)=-{24\Lambda^4\over\pi^4\lambda}\ln\lambda
-{3\over256\pi}{M^2\over G\lambda^3},
\label{de2}
\end{equation}
where the last term stems from the binding energy between pairs of wormholes.
The result of the minimization is 
\begin{equation}
\lambda\simeq e,
\label{l}
\end{equation}
and
\begin{equation}
M=G^{-{1\over2}}=m_p,
\label{mp2}
\end{equation}
the extreme of the allowed range for the ADM-mass $M$ (see eq.(\ref{mp1})).

The above chain of deductions delivers us a state of the quantized
gravitational field whose energy density (as seen by an asymptotic observer
living over a flat space-time) is vastly $[O(\Lambda^4)]$ below the energy
density of the Perturbative Ground State. This state consists of a
(semi)-classical background whose structure is that of an interacting gas of
wormholes whose average distance is of the order of the Planck length
$a_p=G^{1\over2}$, and its average ADM-mass the Planck mass
$m_p=G^{-{1\over2}}$. Recalling that for the outside observer the radius $2GM$
of the wormhole is an unpenetrable horizon, the space-configuration of this
putative GS is equivalent to a random lattice with average lattice constant
$a_p$. Indeed the only reasonable and meaningful description our observer can
make of this peculiar configuration is to assign an average value of the metric
field to the spatial interstices existing among 
the voids created by the wormholes. An
adequate approximate picture of space-time that captures the essential features
of such state is then that of a 
regular lattice with the lattice constant $a_p$:
the Planck lattice. 

\section{\bf Conclusion}

In this letter we have tried to gain some understanding of the structure of the
GS of Einstein's Quantum Gravity. In order to achieve this ambitious goal we
have focussed our attention on the 
gauge-structure of Einstein's General Relativity (GR) and
envisaged a strategy of analysis that for a non-abelian gauge theory, such
as QCD, had proved deeply insightful \cite{gp}. We have thus tried out a
variational calculation of the appropriately defined energy functional of the
quantum fluctuations of the gravitational field around the classical solution
of matterless GR: Schwarzschild's wormholes. Having found negative eigenvalues
(associated to ``unstable modes'') for the second order wave-operator arising
from the lowest order expansion of the GR Hamiltonian, we have given arguments
for the cancellation of the classical $[O(1)]$ term of the energy, receiving 
contribution from the wormhole and 
the  $S$-wave ``unstable mode''. Deprived of the 
positive classical energy term the energy 
density of the quantum gravitational field around a wormhole turns out to be
lower than that of the quantum fluctuations over flat space-time, the 
Perturbative Ground State, thus explicitly proving that the PGS of
QG and the awfully non-renormalizable perturbation theory 
are physically meaningless,
due to the ``essential'' \cite{gp} instability of the PGS.

By extending the single wormhole configuration to a multi-wormhole one, 
we have discovered that
the size $2GM$ of the wormholes is quantum mechanically limited by the
Planck length $a_p=G^{1\over2}$ and, consequently,
the ADM-mass by the Planck mass $m_p=G^{-{1\over2}}$. As a result in the
the state of minimum energy space is seen to acquire the structure of a
random lattice with average lattice constant $a_p$. A space-structure that can
be adequately modeled by a Planck lattice.

Before discussing the physics conclusions that one may draw from such 
a characteristic structure of the putative GS of QG, we would like to warn the 
reader about the still speculative, though eminently plausible nature of our
results.
The only thing of this work that definitely stands 
on solid ground, in fact, is the 
discovery of the instability of 
the PGS of Quantum Gravity, and of its unavoidable
irrelevance for the quantum dynamics of the gravitational field. We deem this 
fact quite important, for it clears the way of QG from the embarassments of 
a disastrously unrenormalizable theoretical structure, that only belongs, 
as far as we know, to an approach, PT, whose very foundations are 
thus seen to dematerialize.  
Having so disposed of the {\it pars destruens}, what can we
say about the {\it pars construens} of this paper?

If we accept that wormholes are an essential ingredient not of a local 
minimum of the energy density in the configurational space
of the quantum gravitational field, but of the absolute minimum, the true
ground state, then a satisfactory solution appears in sight of the long standing
problem of local Quantum Field Theories (QFT's), that so upset among
others Dirac: the infinities that the renormalization program cleverly
manipulates but does not really eliminate from the theory. Indeed,
if the small scale structure of space-time is well described by a Planck 
lattice of lattice constant $a_p$, then for the QFT's of the Standard Model
there exists a natural cutoff at the Planck mass $m_p$, and the logarithmic
infinities of those ``renormalizable''
continuous QFT's become now {\it small corrections}, as appropriate for
terms of a perturbative series. A most welcome result.

We find in this plausible derivation (modulo the {\it caveats} already
expressed) of Wheeler's Quantum Foam, that realizes the prophetical ideas of
Riemann's ``\"Uber die Hypothesen welche der Geometrie zu Grunde liegen''
\cite{re}, the most pleasing and possibly fertile outcome of this work.

\end{document}